\documentclass[11pt]{article} 

\usepackage[ansinew]{inputenc}
\usepackage{amsmath, amssymb, graphics, amsthm}
\usepackage{epsfig}
\usepackage{color}
\usepackage{fancyhdr} 

\usepackage[normalem]{ulem}

\newcommand{\ket}[1]{\left| #1 \right\rangle}

\makeatletter
\@addtoreset{equation}{section}
\makeatother

\newcommand{\be}{\begin{equation}}
\newcommand{\ee}{\end{equation}}
\newcommand{\ba}{\begin{eqnarray}}
\newcommand{\ea}{\end{eqnarray}}

\def\pb#1{\rlap{\lower1.5ex\hbox{$\longleftarrow$}}{#1}}
\def\dpb#1{\rlap{\lower1.5ex\hbox{$\Longleftarrow$}}{#1}}
\def\spb#1{\rlap{\lower1.0ex\hbox{$\leftarrow$}}{#1}}
\def\sdpb#1{\rlap{\lower1.0ex\hbox{$\Leftarrow$}}{#1}}

\title{{\sf A note on entanglement entropy and quantum geometry}} 
\author{
{\sf N. Bodendorfer}\thanks{{\sf 
norbert.bodendorfer@fuw.edu.pl}}\\
{\sf  Faculty of Physics, University of Warsaw,}
{\sf   Pasteura 5, 02-093, Warsaw, Poland}\\
}
\date{{\small\sf \today}}

\begin{document} 

\maketitle

{\sf

\begin{abstract}
It has been argued that the entropy which one is computing in the isolated horizon framework of loop quantum gravity is closely related to the entanglement entropy of the gravitational field and that the calculation performed is not restricted to horizons. 
We recall existing work on this issue and explain how recent work on generalising these computations to arbitrary spacetime dimensions $D+1 \geq 3$ supports this point of view and makes the duality between entanglement entropy and the entropy computed from counting boundary states manifest. In a certain semiclassical regime in $3+1$ dimensions, this entropy is given by the Bekenstein-Hawking formula. 
\end{abstract}

}

\section{Introduction}

Despite remarkable success in the computation of black hole entropy from many approaches to quantum gravity, the deeper meaning of the thermodynamic properties of black holes remains obscure \cite{WaldTheThermodynamicsOf}. While the individual approaches generally agree on the answer to the question ``How large is the entropy of a black hole in a given gravitational theory?'' by stating its classical Wald entropy (plus possible quantum corrections), their means to arrive at this answer are very different. Some examples are entanglement entropy across the horizon \cite{BombelliQuantumSourceOf, SrednickiEntropyAndArea}, the entropy derived from D-branes in string theory \cite{StromingerMicroscopicOriginOf}, and the entropy via quantum geometry from loop quantum gravity (LQG) \cite{KrasnovGeometricalEntropyFrom, RovelliBlackHoleEntropy}. 
It is thus imperative to obtain a thorough understanding of the relation between these different derivations. 

A starting point is provided by realising that the black hole entropy derived from loop quantum gravity is essentially given by the entanglement entropy of the gravitational field across the horizon in a certain state. 
The evolution of this argument actually has a rather long history and is rooted in the concept of edge (=boundary) states in a gauge theory: whenever a holonomy of a gauge field terminates on the boundary (=edge) of the manifold we are considering, one needs to introduce an edge state transforming under gauge transformation such that when contracted with the holonomy, the whole state is gauge invariant. Most famously, these edge states were used in the explanation of the quantum Hall effect \cite{HalperinQuantizedHallConductance}. The general concept was discussed more thoroughly in \cite{BalachandranEdgeStatesIn} with an emphasis on $1+1$ and $2+1$ dimensions. The importance of edge states for general relativity was emphasised in \cite{BalachandranEdgeStatesInGravity} and their relation to entanglement entropy was proposed in \cite{BalachandranEdgeStatesAnd}.

Within loop quantum gravity, such a relation has been advocated first by Husain \cite{HusainApparentHorizonsBlack}. He realised that the entropy computation performed via boundary state counting of quantum geometries in the isolated horizon framework \cite{AshtekarQuantumGeometryOf} was insensitive to many details of the original proposal, e.g. the boundary being a horizon or the dynamics of the theory even admitting black holes.
In fact, Smolin's seminal paper \cite{SmolinLinkingTopologicalQuantum}, which, inspired by the work of Crane \cite{CraneClockAndCategory}, provided the basis for the entropy computations within loop quantum gravity, was already not restricted to horizons. Also, Krasnov's proposal to associate a geometric entropy to a surface \cite{KrasnovGeometricalEntropyFrom} was not. However, in order to make some of the ideas of \cite{SmolinLinkingTopologicalQuantum, KrasnovGeometricalEntropyFrom, RovelliBlackHoleEntropy} precise and to have a better physical motivation for the computation, \cite{AshtekarQuantumGeometryOf} restricted the boundary to an isolated horizon.

Given the observation in e.g. \cite{HusainApparentHorizonsBlack} that the entropy computed from LQG should be identified with the entanglement entropy across an entangling surface, it suggests itself to try to extend the results of \cite{AshtekarQuantumGeometryOf} to general boundaries. This will be one of the concerns of this paper. Then however, a good understanding of the entanglement entropy of gauge fields in a lattice-type representation is called for to compare the results to.
Donnelly computed the entanglement entropy of individual spin networks in \cite{DonnellyEntanglementEntropyIn} and that of general states in lattice gauge theories in \cite{DonnellyDecompositionOfEntanglement}. He further showed in \cite{DonnellyEntanglementEntropyIn} that the entanglement entropy agrees with the entropy of the isolated horizon framework \cite{AshtekarQuantumGeometryOf} up to the spin projection (global gauge invariance) constraint discussed in section \ref{sec:FurtherConstraints} and gave general arguments to explain this. 
It should be noted however that the question of entanglement entropy of gauge fields is rather involved due to the gauge redundancy present. Questions arising here are concerned with what the physical degrees of freedom are once a system is divided into two parts (due to gauge non-invariance of open holonomies) and how to precisely deal with the gauge redundancy, see e.g. \cite{BuividovichEntanglementEntropyIn, DonnellyDoGaugeFields, CasiniRemarksOnEntanglement} and references therein. We will encounter this problem later in section \ref{sec:FurtherConstraints} and argue for which seems to be the correct answer in our context. More recently, new progress in computing black hole entropy from an entanglement entropy perspective was made by Bianchi using low energy perturbations \cite{BianchiBlackHoleEntropy} and the boost Hamiltonian of spin foams \cite{BianchiEntropyOfNon}.
See also \cite{ChirocoSpacetimeThermodynamicsWithout} for a discussion in the context of emergent gravity and \cite{DasguptaSemiclassicalQuantizationOf} for a computation of black hole entropy via entanglement entropy from a coherent state peaked on a Schwarzschild black hole. 

The purpose of this paper is to highlight several points concerning the entropy calculation within the isolated horizon framework of loop quantum gravity and its relation to entanglement entropy, which have not received the required attention in the previous literature. On the one hand, we show that the duality between computing entanglement entropy and counting boundary states becomes manifest when considering individual spin networks in the dimension-independent generalisation \cite{BTTXII, BI} of the computation in \cite{AshtekarQuantumGeometryOf}. On the other hand, we emphasise that both computations are valid for general boundaries, leading to the conclusion that the concept of associating an entropy to the boundary of a given region should not be restricted to horizons. Moreover, using the results of \cite{FroddenBlackHoleEntropy, BNI} in $3+1$ dimensions, this entropy is given by the Bekenstein-Hawking formula in a certain semiclassical regime. 
While this paper doesn't provide (long) technical calculations as e.g. its companions \cite{BTTXII, BI, BNII}, it focuses on some conceptual questions left open there which have, to the best of the author's knowledge, not been addressed in a comprehensive fashion elsewhere in the literature.

This paper is organised as follows. In section \ref{sec:EEandBoundary}, we describe the duality between the computation of entanglement entropy and the the counting of boundary states. Next, in section \ref{sec:FurtherConstraints}, we comment on constraints which need to be imposed in the entropy computation when not using Chern-Simons type variables on the boundary. In section \ref{sec:GeneralBoundaries}, we argue that the computations performed so far in the isolated horizon framework are not restricted to horizons, but are valid for general boundaries. The value of the entropy in a certain semiclassical regime is recalled in section \ref{sec:Entropy}. After several comments in section \ref{sec:Comments}, we conclude in section \ref{sec:Conclusion}.

\section{Entropy from entanglement and boundary Hilbert spaces}
\label{sec:EEandBoundary}

The connection between the computation of entanglement entropy and the computation of (black hole) entropy from counting boundary states becomes very clear when considering the dimension independent treatment given in \cite{BTTXII, BI, BNII}. We focus first on the computation of entanglement entropy from a general spin network as given in \cite{DonnellyEntanglementEntropyIn} for the case of SU$(2)$ as a gauge group. To this end, we choose some connected closed region $\Omega$ on the spatial slice on which the spin network is defined, whose boundary $\partial \Omega$ intersects $N$ of the edges of the spin network transversally. These edges then contribute to the entanglement entropy. We neglect the other (non-generic) cases of tangential edges and vertices on $\partial \Omega$, see \cite{DonnellyEntanglementEntropyIn} for details.

It was shown in \cite{DonnellyEntanglementEntropyIn} that the entanglement entropy $S_{\text{EE}}$ associated with such a choice of spin network and region is given by the logarithm of the product of the dimensions of the representation spaces of the spins $j_i$ carried by the intersecting edges, 
\be
	S_{\text{EE}} (\Omega) = \log \prod_{i=1}^N (2j_i+1) \text{.}
\ee
A way to visualise this calculation is to consider the spin network and insert at the intersection points with $\partial \Omega$ a trivial intertwiner, that is the unit matrix in the corresponding representation space. This splits the spin network into two parts inside and outside of $\Omega$. 
The correlation between $\Omega$ and its complement coming from a single edge is captured by the rank of the trivial intertwiner, that is the dimension of the representation space. 
This entropy agrees with the one calculated from the isolated horizon computations \cite{AshtekarQuantumGeometryOf, EngleBlackHoleEntropy} up to corrections resulting from gauge invariance constraints which we will discuss in section \ref{sec:FurtherConstraints}. 

Next to the formulation of loop quantum gravity in terms of Ashtekar-Barbero variables \cite{AshtekarNewVariablesFor, BarberoRealAshtekarVariables} and the corresponding gauge group SU$(2)$, there exists an alternative formulation valid in any spacetime dimension $D+1\geq 3$ in terms of the gauge group SO$(D+1)$, see \cite{BTTVIII} for an overview. This formulation can also use SO$(1,D)$ as the internal gauge group, however the compact group SO$(D+1)$ is preferred for quantisation purposes for both the Lorentzian and Euclidean theory. In addition to the usual Hamiltonian, spatial diffeomorphism and Gau{\ss} constraints, the theory is subject to the simplicity constraints for $D +1\geq 4$, which translate into a restriction on the group representations at the quantum level. It turns out that the allowed ``simple'' representations\footnote{In the mathematical literature, these representations are called most degenerate, (completely) symmetric, class one, or spherical. The notion ``simple'' in this context results from their relation to ``simple'' bi-vectors, that is the product of two vectors \cite{FreidelBFDescriptionOf}.} are labelled by a single non-negative integer $\lambda$ \cite{FreidelBFDescriptionOf} and that in four dimensions, the natural mapping between the two formulations at the level of the Hilbert space (not necessarily the algebra of observables), is given by $\lambda = 2j$ \cite{BTTV}. 
The extension of the entanglement entropy result to higher dimensions is straight forward by just substituting $(2j_i+1)$ by the dimension $d^{D+1}_{\lambda_i}$ of the corresponding simple SO$(D+1)$ representations of the puncturing holonomy. 

A derivation of black hole entropy based on boundary variables in general dimensions in the spirit of \cite{AshtekarQuantumGeometryOf} has been given in \cite{BI}. Here, the boundary variables are densitised internal bi-normals $L^{IJ} = 2 / \beta\, n^{[I} \tilde s^{J]}$, subject to the Poisson bracket 
\be
		\{ L^{IJ}(x), L^{KL}(y) \}  = 4 \,   \delta^{(D-1)}(x-y) \delta^{L][I} L^{J][K}(x)  \text{.} \label{eq:LieAlgebra}
\ee
$n^I$ is the internal analogue of the normal on the spatial slice $\Sigma$ and $\tilde s^I$ is the internal normal on the boundary slice $\partial \Sigma$, densitised with the area density of $\partial \Sigma$ as well as subject to $n^I \tilde s_I = 0$. $\beta$ is a free parameter of the theory, analogous but different from the Barbero-Immirzi parameter. The $L_{IJ}(x)$ have to be regulated by smearing them over small surfaces in the same way as the fluxes in loop quantum gravity, see e.g. \cite{ThiemannModernCanonicalQuantum}. In fact, the boundary condition relating bulk and boundary fields just reads
\be
	\hat s_a \pi^{aIJ} = L^{IJ} {\text{,}} \label{eq:BoundaryCondition}
\ee
where $\pi^{aIJ}$ is the momentum conjugate to the SO$(D+1)$ connection in the bulk and $\hat s_a$ is a properly densitised co-normal on $\partial \Sigma$, indicating that the corresponding flux is integrated over a spacetime-codimension two surface on $\partial \Sigma$. 
A quantisation of a properly regularised form of \eqref{eq:LieAlgebra} now simply yields a non-trivial representation space of SO$(D+1)$ at every point where the bulk spin network punctures $\partial \Sigma$. By the boundary condition \eqref{eq:BoundaryCondition}, these representations are simple. 

Neglecting possible further restrictions of the boundary Hilbert space for the moment, we can now compare this result to the entanglement entropy calculation. The logarithm of the dimension of the boundary Hilbert space is simply given by
\be
	S_{\text{BH}} (\partial \Sigma) = \log \prod_{i=1}^N d^{D+1}_{\lambda_i} \text{,}
\ee
and thus agreeing with the result of the entanglement entropy calculation for $\partial \Omega = \partial \Sigma$ and the bulk state in the second calculation being the restriction of the first spin network in $\Sigma \cup \Omega$ to $\Sigma$. This is traced back to the fact that both computations in the end compute the dimension of the same SO$(D+1)$ representation spaces, in the entanglement entropy picture by ``cutting open'' the holonomies crossing $\partial \Omega$, and in the boundary Hilbert space picture by having exactly these representation spaces induced at points where holonomies puncture $\partial \Sigma$. Morally speaking, introducing the boundary before (counting boundary states) and after quantisation (computing entanglement entropy) commutes for the entropy calculation. See also the general arguments in \cite{DonnellyEntanglementEntropyIn}. The counting of boundary states using a Chern-Simons treatment in $3+1$ dimensions \cite{AshtekarQuantumGeometryOf, EngleBlackHoleEntropy} however yields additional constraints on the boundary Hilbert space, which we will discuss in the next section. We will also discuss the spatial diffeomorphism and Hamiltonian constraints in section \ref{sec:GeneralBoundaries}.

\section{Further constraints from gauge invariance}
\label{sec:FurtherConstraints}

A comparison of the dimension independent results with the SU$(2)$ Chern-Simons treatment  \cite{EngleBlackHoleEntropy} indicates that one might be missing a further constraint which selects a globally gauge invariant subspace in the boundary Hilbert space. 
A way to see that such a further constraint is necessary is to consider the gauge invariant boundary observables $L_i^{IJ} L_{i\,IJ}$, that is the areas, in the neighbourhood of a puncture $i$. Since the SO$(D+1)$ gauge transformations generated by the Gau{\ss} constraint
\be
	G^{IJ}[\Lambda_{IJ}] = \int_\Sigma d^Dx\,  \Lambda_{IJ} D_a \pi^{aIJ} = - \int_\Sigma d^Dx\,  (D_a \Lambda_{IJ}) \pi^{aIJ} + \int_{\partial \Sigma} d^{D-1}x\,  \Lambda_{IJ} L^{IJ}
\ee
act locally a priori \cite{BI}, there are no further gauge invariant and independent boundary observables in terms of the bi-normals\footnote{We note that constructing a connection from the bi-normals after quantisation to emulate the Chern-Simons treatment is not feasible due to mathematical difficulties.}. $D_a$ here denotes the internal covariant derivate with respect to the SO$(D+1)$ connection conjugate to $\pi^{aIJ}$. What is problematic is that the representation of these boundary operators is not irreducible, since they act diagonally on the individual SO$(D+1)$ representation spaces. In other words, the boundary Hilbert space is too big and one risks an overcounting.  Experience with the SU$(2)$ Chern-Simons theory now suggests to restrict to a global gauge invariance acting only on the boundary observables, which at the same time restricts the boundary Hilbert space and allows for more gauge-invariant operators. This is because the local gauge transformations can be compensated in the Chern-Simons theory, which is defined in terms of a connection already classically, by considering holonomies running between punctures.
For the bi-normals, this effectively corresponds to introducing the additional constraint 
\be
	\sum_i \hat{L}_i^{IJ} \ket{\Psi_{\text{Boundary}}} = 0 \label{eq:GlobalGauge}
\ee
and restricting the Lagrange multiplier of the Gau{\ss} constraint to be constant on $\partial \Sigma$ \cite{BI}.
The additional gauge invariant operators now contain the ``angles'' 
\be
	\frac{L_i^{IJ} L_{j\,IJ}}{  \sqrt{L_i^2} \sqrt{L_{j}^2}} \text{.} \label{eq:Angles}
\ee
From a Chern-Simons point of view, such a restriction would not be necessary, but it would be effectively enforced by considering gauge invariant states, that is contractions of holonomies on the boundary with intertwiners. Also, the dimension of the boundary Hilbert space then agrees with the dimension of the intertwiner space of a (hypothetical) intertwiner inside the boundary which contracts all incoming holonomies (up to the off-diagonal simplicity constraints discussed next).

It was further argued in \cite{BI} in the context of a spherical non-rotating isolated horizon that one also needs to impose (a maximally commuting subset \cite{BTTV} of) the constraints
\be
	\hat L_i^{[IJ} \hat L_{j}^{KL]} \ket{\Psi_{\text{Boundary}}}= 0 \label{eq:OffSimplicity}
\ee
similar to the off-diagonal simplicity constraint for $i \neq j$ and diagonal simplicity constraints for $i=j$ \cite{FreidelBFDescriptionOf}. Such constraints restrict the intertwining representations in a certain recoupling scheme to be simple, leading to a unitary equivalence of the boundary Hilbert spaces in any dimension \cite{BTTV, BI} for a given set of punctures with labels $\lambda_i$. This way, a maximally commuting, that is simultaneously diagonalisable, subset of the angles \eqref{eq:Angles} are the only new non-trivial gauge invariant operators and one again obtains an agreement with the picture of having an allowed intertwiner of the theory \cite{BTTV} inside the boundary. 
Structurally, this angle subset is built in the same way as the subset for the off-diagonal simplicity constraints, that is it reads out the representation labels of the simple representations in the chosen recoupling scheme. We note that the maximally commuting subset of angles are weak Dirac observables with respect to the maximally commuting subset of simplicity constraints. 
As noted in \cite{BI}, the number of boundary states is independent of the choice of maximal subset.
The necessity for these additional constraints comes from the fact that the $L_i^{IJ}$ are a redundant description of the physical system. The linear simplicity constraints, which are already enforced by the boundary condition \eqref{eq:BoundaryCondition}, tell us that all the  $L_i^{IJ}$ factorise as $L_i^{IJ}(x) = n^{[I}(x) \tilde s^{J]}(x)$. However, we also need to impose that the choice of $n^I(x)$ is pure gauge, which can e.g. be done by introducing the additional constraint that the $n^I(x)$ should agree. This is enforced by the off-diagonal simplicity constraints and the generator of global gauge transformations \eqref{eq:GlobalGauge}, which mods out remaining global gauge transformations on the $n^I(x) = n^I(y)$ after imposing off-diagonal simplicity.
We leave the question of possible topological obstructions for the gauge fixing in non-spherical boundary topologies open for now. Such obstructions might be related to topology corrections to the entropy as anticipated in \cite{BI}.

We thus adopt these constraints also for general boundaries, noting again that this leads to an agreement of the quantisations in different variables. Eventually, one would like to deduce these constraints from a rigorous treatment of the boundary theory as a higher-dimensional Chern-Simons theory, see e.g. \cite{BTTXII} and references therein. The argument in $3+1$ dimensions could run as follows: we know from the classical derivation that we are not quantising a generic SO$(4)$ Chern-Simons theory, but one in terms of a specific SO$(4)$ connection $\Gamma^0$ subject to certain constraints \cite{BTTXII}. One of them is that its curvature $R^0$ satisfies $R^0_{\alpha \beta IJ} n^I = 0$. $\alpha, \beta$ are local indices on a spatial slice $\partial \Sigma$ of the boundary and $I, J, \ldots$ SO(4) indices. It now follows that $\epsilon^{IJKL}R^0_{\alpha \beta IJ} (x)R^0_{\gamma \delta KL}(x) = 0$, corresponding to the diagonal simplicity constraints. In order to map between the internal spaces at two points $x$ and $y$ on the boundary, we can use the parallel transporter $U_\gamma^0(x,y)$ along a path $\gamma$ constructed from $\Gamma^0$. Then, $\epsilon^{IJKL}R^0_{\alpha \beta IJ} (x) U_\gamma^0(x,y)_K {}^M  U_\gamma^0(x,y)_L{}^N R^0_{\gamma \delta MN}(y) = 0$ since the expression is gauge invariant and it vanishes choosing the gauge $n^I = (1,0,0,0)$ in a neighbourhood of $\gamma$. This results from the fact that in this gauge $\Gamma^0_{\alpha IJ} n^J = 0$. This argument generalises to higher dimensions by taking the appropriate wedge products of $R^0$, see \cite{BTTXII}. An actual quantisation of non-abelian higher-dimensional Chern-Simons theory is however not available at the moment and complicated by local degrees of freedom. 
Alternatively, one could try to generalise the approach of \cite{MaTheBFTheory} without gauge fixing the boundary theory.

Another way to gain some insight in this problem is related to the well-known second class nature of the off-diagonal simplicity constraints. 
We first note that the action of the generator of local gauge transformations on the states we are considering vanishes already, since the action of its quantisation coincides with that of the boundary condition \eqref{eq:BoundaryCondition} and is solved simply by contracting puncturing holonomies with the boundary representation spaces \cite{BI}. We thus should focus on how it selects boundary observables. As noted before, the areas associated to the punctures are the only gauge invariant non-trivial boundary observables in terms of the bi-normals, which means that for the purpose of computing the entropy, we should count only a single state for a distinct set of areas. 
We now adopt a gauge unfixing \cite{MitraGaugeInvariantReformulationAnomalous} point of view and impose as gauge fixings the generator \eqref{eq:GlobalGauge} of global gauge transformations and all of the off-diagonal simplicity constraints \eqref{eq:OffSimplicity}. The simplicity constraints now enforce that there is a common normal $n^I$ shared by all the $L_i^{IJ}$, whereas the remaining gauge transformations are (on-shell) generated by diagonal simplicity constraints (orthogonal to the $n-s$-direction), the off-diagonal simplicity constraints (in the $s - \perp(n-s)$-direction), and the generator of global gauge transformations \eqref{eq:GlobalGauge} (global transformations). Local gauge transformations in the $n-s$ planes act trivially on the $L_i^{IJ}$. Note that by the vanishing of \eqref{eq:GlobalGauge}, the remaining gauge transformations (potentially violating \eqref{eq:GlobalGauge}) are gauge fixed. One could now argue to gauge unfix the system by dropping the generator of local gauge transformations. However, the problem with this strategy is that the off-diagonal simplicity constraints do not form a closing algebra, see e.g. \cite{BTTIII}, and thus cannot be interpreted as the generators of gauge transformations. In fact, imposing all of them strongly leads to a single allowed boundary state, the higher-dimensional analogue of the Barrett-Crane intertwiner \cite{FreidelBFDescriptionOf}. The only observables commuting with all the constraints are then the areas associated to the punctures. We are thus in the same situation as before. 
To avoid this problem, one can resort to a similar strategy as for the off-diagonal simplicity constraints in the bulk acting on spin network vertices \cite{BTTV}. One only imposes a maximally commuting subset \cite{BTTV} of the off-diagonal simplicity constraints, which has a maximally commuting subset of the angles \eqref{eq:Angles} as observables and generates (or gauge fixes) only gauge transformations which leave this subset of \eqref{eq:Angles} invariant. 
In conclusion, this procedure leads to the same result as the arguments in \cite{BI} and mimics the strategy employed in the bulk \cite{BTTV}. 
The problem with employing this strategy is of course that we are not performing a proper gauge unfixing, since taking only a maximally commuting subset of simplicity constraints doesn't gauge fix all local rotations. Still, it provides some insight in the issue, as it relates the occurring problems to the well-known problems with the (higher-dimensional analogue of the) Barrett-Crane intertwiner.

Collecting our results, it seems to be the proper procedure to impose a global gauge invariance condition on the representation spaces arising in the computation of the entanglement entropy. This results from the fact that we are dealing with a gauge theory where the physical (gauge invariant) states are non-local. Furthermore, a maximally commuting subset of the off-diagonal simplicity constraints has to be imposed to deal with further gauge redundancy of the theory. These constraints were argued to arise naturally in a Chern-Simons type quantisation of the boundary theory and reflect the gauge redundancy present.

\section{General boundaries}
\label{sec:GeneralBoundaries}

The classical parts of the derivation of the entropy computations in the isolated horizon framework \cite{AshtekarIsolatedHorizonsThe, AsthekarQuantumHorizonsAnd, BeetleGenericIsolatedHorizons, EngleBlackHoleEntropyFrom, BTTXII} were based on certain isolated horizon conditions imposed on null boundaries. The motivation for this line of reasoning was clearly motivated by the goal to reproduce the Bekenstein-Hawking entropy of black holes. Despite the fact that it was quickly realised that it was actually inessential for the boundary to be null \cite{HusainApparentHorizonsBlack}, the focus of the computations performed remained on horizons. Arguably, this was largely due to the fact that the calculations performed focused on pulling back spacetime connections to the horizon. Upon using the isolated horizon boundary conditions, these connections then reduced to the (gauge fixed) ones that one would have obtained form a treatment for general boundaries. 
A $3+1$-dimensional construction where the connection on the horizon was not based on the induced metric on the horizon was first given in \cite{AsthekarQuantumHorizonsAnd} for axially symmetric horizons and generalised in \cite{BeetleGenericIsolatedHorizons} to arbitrary isolated horizons. A similar construction works in general dimensions \cite{BNII}. The essential idea is that the connection on the boundary doesn't need to be based on the actual induced metric on the boundary, but only on a metric sharing the same area element. This metric can then be tuned so that the associated connection satisfies the proper boundary condition \eqref{eq:FGS}. 

The essential ingredients for the entropy computation in the isolated horizon framework are the boundary condition
\be
	\hat s_a \pi^{aIJ} \propto \epsilon^{IJKL} \epsilon^{\alpha \beta} \, F_{\alpha \beta KL}(\Gamma^0) \label{eq:FGS}
\ee
 relating bulk and boundary degrees of freedom, as well as the boundary symplectic structure
 \be
 	\Omega_{\text{Boundary}}(\delta_1, \delta_2) \propto	\int_{\partial \Sigma} d^2x \,	 \epsilon^{IJKL} \epsilon^{\alpha \beta} \, \delta_{[1} \Gamma^0_{\alpha IJ} \,  \delta_{2]} \Gamma^0_{\beta KL} \text{.} \label{eq:BSS}
 \ee 
We presented here the case of $3+1$ dimensions with internal gauge group SO$(1,3)$ in Chern-Simons form given in \cite{BTTXII} for definiteness, however the following statements are also true for the bi-normal form presented in section \ref{sec:EEandBoundary}, in higher dimenions (even spacetime dimensions in case of the Chern-Simons form), and for SO$(D+1)$ as gauge group. $\alpha, \beta$ are local indices on a spatial slice $\partial \Sigma$ of the boundary and $I, J, \ldots$ SO(1,3) indices. $\Gamma^0_{\alpha IJ}$ is an SO$(1,3)$ connection on $\partial \Sigma$ defined in \cite{BTTXII} and $F$ its curvature.
The main point of this section is to highlight that \eqref{eq:FGS} and \eqref{eq:BSS} are not restricted to isolated horizon boundaries.  In fact, the dimension-independent derivation in \cite{BTTXII} of the boundary condition and boundary symplectic structure is, up to the non-distortion condition lifted in \cite{BNII}, completely independent of the type of boundary used. This is of course only provided one does not insist on the boundary connection being the pullback of a spacetime connection in the formulation using SO$(1,3)$ as a gauge group. However, as said before, this is inessential for the calculation. Another focus of \cite{BTTXII} was to show that the variational principle was well defined given isolated horizon boundary conditions. We can however choose a different boundary condition, e.g. $\delta q_{ab} = 0$, with $q_{ab}$ being the induced metric on a general boundary $\partial \Omega$. This would lead to the York-Gibbons-Hawking boundary term of the action, see \cite{BNI} and references therein for our context. While $\delta q_{ab}=0$ on $\partial \Omega$ ensures a well defined variational principle, we can still have non-trivial internal gauge transformations on the boundary when going over to connection variables and thus densitised bi-normals or Chern-Simons variables on the boundary\footnote{The analogous boundary condition using an orthonormal frame is $\delta{(e_a^I e_{bI})}=0$ on $\partial \Omega$, see \cite{BNI} for details.}. Thus, the framework to compute the entropy is still intact. In other words, even fixing the induced metric on the boundary, both the boundary condition and boundary symplectic structure work as before. 
We are thus lead to associate the same entropy to null and general boundaries\footnote{Again, more precisely the entropy is associated to the boundary of a certain spatial slice (intersecting the spacetime boundary). Demanding that the area of these slices' boundaries stays fixed throughout the time evolution, the entropy would be independent of the chosen slice, as it is e.g. in the isolated horizon framework.}.

We now need to discuss the remaining constraints of the theory. As remarked before, our aim is to compute the entropy associated to the boundary $\partial \Sigma$ of a certain spatial slice $\Sigma$. This means that the gauge transformations generated by the constraints of the theory have to be restricted such that they preserve $\partial \Sigma$. In short, we need to demand that $\hat s_a N^a = 0$ on $\partial \Sigma$ for the shift vector and $N = 0$ on $\partial \Sigma$ for the lapse function used to smear the spatial diffeomorphism and Hamlitonian constraints. 
The canonical analysis of the general relativity action with York-Gibbons-Hawking boundary term on general boundaries has been given in \cite{HawkingTheGravitationalHamiltonian}. The spatial diffeomorphism constraint is given by 
\be
	\mathcal{H}_a[N^a]~ = ~ \int_\Sigma d^Dx\, P^{ab} \mathcal{L}_N q_{ab} ~ \approx ~  \frac{1}{2} \int_\Sigma d^Dx\, \pi^{aIJ} \mathcal{L}_N A_{aIJ} + \frac{1}{\beta} \int_{\partial \Sigma} d^{D-1}x\,  n^I \mathcal{L}_N \tilde s_I \text{,} \label{eq:DiffConstraint}
\ee
where $\mathcal L_N$ denotes the Lie derivative with respect to the shift vector $N^a$\footnote{In \cite{HawkingTheGravitationalHamiltonian}, this constraint is split into $N_a D_b P^{ab}$ and a boundary term denoted ``momentum term''. We include this boundary term in the constraint for it to generate spatial diffeomorphisms also at the boundary. The classical expression for the entropy is invariant under such diffeomorphisms, which is why it makes physical sense to mod them out in the quantum theory.}. $\approx$ here means equality up to a term proportional to the boost part $n_I G^{IJ}$ of the Gau{\ss} constraint. The spatial diffeomorphism constraint thus generates spatial diffeomorphisms on both the bulk and boundary variables and a quantisation of it selects diffeomorphism equivalence classes of spin networks (possibly puncturing $\partial \Sigma$), see e.g. \cite{ThiemannModernCanonicalQuantum} for details. 
The Hamiltonian constraint obtains a boundary contribution \cite{HawkingTheGravitationalHamiltonian}, however as said before, we demand $N=0$ on $\partial \Sigma$ for the Hamiltonian constraint to preserve the chosen spatial boundary. In the quantum theory, we assume that for every boundary state, there exists a compatible bulk state in the kernel of the Hamiltonian constraint, as is usually done \cite{AsthekarQuantumHorizonsAnd}. 
Also an explicit treatment of the Hamiltonian constraint in the context of computing entanglement entropy leads to a similar conclusion: in the current regularisation of the constraint \cite{ThiemannQSD1}, it acts only on spin network vertices, and not on edges, e.g. the ones crossing the boundary. Another approach would be to solve the Hamiltonian constraint classically by gauge fixing, leading to a similar conclusion \cite{GieselAQG4, BSTI}. One could e.g. define the spatial boundary of our choice by having a scalar field assume a certain value on it.

\section{The entropy in a semiclassical regime}
\label{sec:Entropy}

What remains to be done is to compute the value of the entropy. The main conceptual issue here is that one would expect the spin network states we are considering here to be highly quantum states with little semiclassical interpretation. More precisely, we would need to compare the entropy with the effective action of the theory derived from a path integral. In $3+1$ dimensions, this is actually possible for large quantum numbers (spins), that is large area eigenvalues. The asymptotic analysis of the corresponding EPRL-FK spin foam model \cite{FreidelANewSpin, EngleLQGVertexWith} for a flat $4$-simplex in this limit has been given in \cite{BarrettLorentzianSpinFoam}. Inspired by the results of \cite{FroddenBlackHoleEntropy}, it was realised in \cite{BNI} that not only the real part of the general relativity action is reproduced correctly, but also its imaginary part \cite{NeimanOnShellActions, NeimanTheImaginaryPart, NeimanAsymptotic} when performing an analytic continuation to a complex Barbero-Immirzi parameter $\gamma = \pm i$. By the result of \cite{FroddenBlackHoleEntropy}, the entropy becomes $A / 4G$ in the same limit. Also this reasoning is not restricted to horizons. In fact, the focus of $\cite{BNI}$ was to consider quantum gravity in finite regions, where the analysis of \cite{BarrettLorentzianSpinFoam} is most relevant.
Thus, we have found that based on the computations performed so far within loop quantum gravity, one should associate an entropy of $A/4G$ to a general boundary in the large spin semiclassical regime (see the following comment for a disclaimer on the terminology ``semiclassical'').

\section{Comments}
\label{sec:Comments}

\begin{itemize}

	\item While the discussion in this paper has been in the context of dimension-independent connection variables, it also applies also to the SU$(2)$-based Ashtekar-Barbero variables \cite{AshtekarNewVariablesFor, BarberoRealAshtekarVariables}. In fact, it was shown in \cite{BTTXII} that in $3+1$ dimensions, the boundary symplectic structure can be rewritten to coincide with the one of Chern-Simons theory in terms of a two-parameter family of connection variables. Choosing time gauge and removing the dimension-independent part of the connection leads to Ashtekar-Barbero variables. The sole restriction imposed in \cite{BTTXII} for this computation, the non-distortion condition, has been lifted in \cite{BNII}, so that the boundary symplectic structure for any boundary in $3+1$ dimensions can be massaged to be of SU$(2)$ Chern-Simons type, along with the respective boundary condition analogous to \eqref{eq:FGS}.
		
	\item There are a priori different (semiclassical) regimes in the theory in which an entropy can be calculated. The original computations \cite{KrasnovGeometricalEntropyFrom, RovelliBlackHoleEntropy, AshtekarQuantumGeometryOf} of  counting all horizon states with given boundary area correspond to computing the entanglement entropy in a maximally mixed boundary state including all possible decompositions of the horizon area into quanta of area. The effective action in this regime of the theory is however not know. On the other hand, more is possible in the large spin limit along with the analytic continuation proposed in \cite{FroddenBlackHoleEntropy}. Here, a comparison to an effective action is possible and yields agreement \cite{BNI}. However, it was argued in \cite{BNI} that this regime of the theory, while being semiclassical in the sense of a stationary path integral, exhibits a ``transplanckian'' character, see e.g. \cite{DvaliPhysicsOfTrans} and references therein.

	\item In the large spin regime in $3+1$ dimensions, there is by now a logically rather coherent picture for the entropy computation, including a comparison with an effective action. In dimensions other than $3+1$ however, many details are still missing. One would also like to perform a comparison to an effective action derived from a (to be defined) spin foam model and one would like to have a better understanding of the generalisation of the results of \cite{FroddenBlackHoleEntropy} to general dimensions.
	Still, as already remarked in \cite{BI}, the boundary Hilbert spaces in different dimensions are isomorphic (up to possible topology corrections). This means that the state counting problem is identical, up to the precise form of the area spectrum. Also, the same results are true for Euclidean gravity, since the signature of spacetime enters the present framework only through the Hamiltonian constraint, see e.g. \cite{BTTI}, and for arbitrary cosmological constant, see e.g. \cite{BNII}.
	
	\item The duality between entanglement and quantum geometry becomes manifest in the framework under consideration. Morally speaking, a certain ``quantum'' of entanglement always comes with a ``quantum'' of geometry and vice versa. This provides further evidence for the conjecture stated in \cite{BianchiOnTheArchitecture} that entanglement entropy is a probe for the architecture of spacetime. Similar observations have been stated elsewhere in the literature more or less explicitly, see e.g. \cite{BianchiBlackHoleEntropy} and references therein.
	
	\item In the context of holography \cite{BoussoTheHolographicPrinciple}, the results of this paper fall into the category of ``weak holography'' in the language of \cite{SmolinTheStrongAnd}. That is, the degrees of freedom measurable at the boundary of a region are bounded by the exponential of the Bekenstein-Hawking entropy (in the discussed semiclassical regime), however no assertion about the actual degrees of freedom within the bounded region is made. ``Strong holography'' only holds if one considers the simplest subset of states in the interior of the bounded region, that is all ingoing holonomies contracted by a single intertwiner, corresponding to a single ``atom of space'' as the bounded region\footnote{Here we also ignore the issue of remaining geometric moduli arising from semi-analytic diffeomorphisms \cite{ThiemannModernCanonicalQuantum}.}. 
See also \cite{JacobsonOnTheNature} for a discussion in the context of the interpretation of black hole entropy and \cite{SchroerBondiMetznerSachs} for a quantum field theory viewpoint. 

	\item The imaginary part of gravity actions has a close connection to entanglement entropy already at the classical level \cite{NeimanActionAndEntanglement}. Thus, it is logically sound that the entropy derived here assumes it expected value $A/4G$ if also the imaginary part of the effective action is reproduced correctly via analytically continuing the Barbero-Immirzi parameter.  

	\item The results of this paper extend to Lanczos-Lovelock gravity by the results of \cite{BNII} when the boundary is restricted to be a non-rotating isolated horizon. The only difference is then that area, e.g. the area elements, will be replaced by the appropriate Wald entropy \cite{JacobsonEntropyOfLovelock, WaldBlackHoleEntropy} elements. We note that this can provide a non-perturbative mechanism for an area-proportional matter field entanglement entropy becoming proportional to the Wald entropy when coupled to quantum Lanczos-Lovelock gravity, since the available ``quantum channels'' across the horizon are now determined by the quantised Wald entropy instead of the quantised area.

	\item In this paper, we considered a boundary that only intersects edges of the spin network transversally. This seems to be the correct physical situation, as such edges contribute solely to the boundary's area, and not e.g. to the volume. Considering also non-trivial vertices on the boundary leads to several problems, one of which is that the entropy computation doesn't work as before as e.g. the precise form of the allowed diffeomorphisms needs to be taken into account, see e.g. \cite{ThiemannModernCanonicalQuantum}. We neglect this case since it doesn't seem to be physically relevant to us. A thorough understanding of these issues would nevertheless be desirable.
	
	\item The inclusion of matter fields and their proper treatment is a largely open issue in the current framework. While one can argue that on an isolated horizon, matter fields are not independent of the gravitational field and thus need not to be taken into account in the counting \cite{AshtekarQuantumGeometryOf}, this point of view is not applicable to general boundaries and also unsatisfactory from an entanglement entropy point of view. While matter degrees of freedom can be coupled to the theory, see \cite{ThiemannModernCanonicalQuantum} for an overview, an asymptotic analysis of a corresponding matter coupled spin foam model to compare with is missing. In any case, it seems interesting to point out that the computation for the gravitational field is just a special case of the computation for a general gauge field, which would naively lead to an entropy proportional to the (bare) gauge invariant electric Yang-Mills charge as defined in \cite{AshtekarIsolatedHorizonsHamiltonian}, of which the area (Wald entropy) is a special case for the gravitational field. However, such a charge can be suppressed as the square root of the horizon area by non-extremality conditions \cite{CorichiEinsteinYangMills}, which might be of interest for the species problem \cite{SolodukhinEntanglementEntropyOf}.

\end{itemize}

\section{Conclusion}
\label{sec:Conclusion}

In this paper, it has been explained that the black hole entropy computations in the isolated horizon framework of loop quantum gravity make sense also for general boundaries. All necessary ingredients, such as the boundary condition relating bulk and boundary variables and the boundary symplectic structure first derived on an isolated horizon, can be generalised to arbitrary boundaries. Furthermore, the computation was shown to be analogous to the entanglement entropy computation performed in \cite{DonnellyEntanglementEntropyIn}. Thus, one may conclude that the loop quantum gravity entropy calculation is just another way to compute the entanglement entropy of the gravitational field, up to the subtlety with the additional constraints discussed in section \ref{sec:FurtherConstraints}. A similar reasoning was already presented in \cite{BalachandranEdgeStatesAnd} in a more general context, see also \cite{HusainApparentHorizonsBlack}. It is interesting to note that in the regime where the corresponding path integral in terms of the EPRL-FK spin foam model is known to be stationary on (a discretised version of) the general relativity action, the entropy derived for arbitrary boundaries is given by the Bekenstein-Hawking entropy. This provides further evidence for a deep connection between entanglement entropy and the geometry of spacetime as e.g. conjectured in \cite{BianchiOnTheArchitecture}.

\section*{Acknowledgements}

This work was supported by a Feodor Lynen Research Fellowship of the Alexander von Humboldt-Foundation. Useful discussions with Alexander Stottmeister, Etera Livine, Yasha Neiman and Antonia Zipfel are gratefully acknowledged, as well as helpful comments on a draft of this paper by Yasha Neiman and Alexander Stottmeister.

\end{document}